\documentclass[aps,reprint,superscriptaddress,showpacs,amsmath]{revtex4-2}
\usepackage{lmodern}
\usepackage{physics}

\usepackage{lmodern}
\usepackage[T1]{fontenc}
\usepackage{color}
\usepackage{amstext}
\usepackage{amssymb}
\usepackage{amsmath}
\usepackage{graphicx}
\usepackage{esint}
\newcommand{\red}[1]{\textcolor{black}{#1}}

\usepackage{setspace}   

\begin{document}
\title{Quantum defects from single surface exhibit strong mutual interactions}

\author{Chih-Chiao Hung} 
\email{chih-chiao.hung@riken.jp}
\affiliation{Laboratory for Physical Sciences, 8050 Greenmead Drive, College Park, Maryland 20740, USA}
\affiliation{Quantum Materials Center, University of Maryland, College Park, Maryland 20742, USA} 
\affiliation{Department of Physics, University of Maryland, College Park, Maryland 20742, USA} 
\author{Tim Kohler} 
\affiliation{Laboratory for Physical Sciences, 8050 Greenmead Drive, College Park, Maryland 20740, USA}
\author{Kevin D. Osborn } 
\email{corresponding author, osborn@lps.umd.edu}
\affiliation{Laboratory for Physical Sciences, 8050 Greenmead Drive, College Park, Maryland 20740, USA}
\affiliation{Quantum Materials Center, University of Maryland, College Park, Maryland 20742, USA}
\affiliation{Joint Quantum Institute, University of Maryland, College Park, MD 20742, USA}

\date{\today}

\begin{abstract}
Two-level system (TLS) defects constitute a major decoherence source of quantum information science, but they are generally less understood at material interfaces than in deposited films.
Here we study surface TLSs at the metal-air interface, by probing them using a quasi-uniform field within vacuum-gap (VG) capacitors of resonators.
The VG capacitor has a nano-gap which creates an order-of-magnitude larger contribution from the metal-air interface than typical resonators used in circuit QED. We measure three phenomena and find qualitative agreement with an interacting TLS model, where near-resonant TLSs experience substantial frequency jitter from the state switching of far-detuned low-frequency TLSs. First, we find that the loss in all of our VG resonators is weakly or logarithmically power dependent, in contrast to data from deposited dielectric films.
Second, we add a saturation tone with power $P_{in}$ to a transmission measurement and obtain the TLS Rabi frequency $\Omega_{0}$.
These data show a substantially weaker $P_{in}$ dependence of $\Omega_{0}$ than the prediction from the standard non-interacting TLS model.
Lastly, we increase the temperature and find an increased TLS jitter rate and dephasing rate from power-dependent loss and phase noise measurements, respectively.
We also anneal samples, which lowers the low-frequency TLS density and jitter rate, but the single-photon loss is found to be unchanged.
The results are qualitatively consistent with a fast-switching interacting-TLS model and they contrast the standard model of TLSs which describes TLSs independently.

\end{abstract}

\maketitle

\section{Introduction}

Suppressing dissipation and noise by choosing materials with a low defect density is essential for superconducting \cite{martinis2005decoherence,gambetta2017building} and semiconducting \cite{Nichol2019} qubits.
These defects are known as two-level systems (TLSs) due to their characteristic two energy levels \cite{STM1,STM2} and cause energy absorption through their dipole moment $p$.
For instance, $\mathrm{SiN_x}$ and $\mathrm{SiO_2}$ host TLSs at a known density, which limits the coherence time of qubits \cite{martinis2005decoherence, paik2010reducing}; presently most qubits do not use these dielectrics but still rely on the AlOx dielectric in the Josephson junction (JJ) barrier.
Additionally, kinetic-inductance photon detectors \cite{MKID2009,MKID2012}, quantum-limited microwave amplifiers \cite{white2015traveling}, and quantum transducers \cite{MechRes2013,MechRes2016} can be affected by TLSs. 
TLSs are located in all known deposited dielectric films \cite{martinis2005decoherence,paik2010reducing}, and material interfaces including surface oxides \cite{wang2015surface,gambetta2016,Oliver2019Res,NbOx2021}.
There are attempts to prevent the growth of the native oxide of metals using vacuum packaging \cite{mergenthaler2021ultrahigh} or to reduce it by HF-treatment \cite{megrant2012planar,verjauw2021investigation}.
Understanding and mitigating TLSs can further improve the quality of low temperature devices.

Measuring the loss  $1/Q_i$ (the inverse of the quality factor) of a resonator as a function of photon number $n_{ph}$ provides a standard way to extract information on TLSs \cite{STM1,STM2}. A standard fitting formula for this purpose is
\begin{equation}
    \frac{1}{Q_i} = \frac{1}{Q_i^0} \frac{\mathrm{tanh}(\frac{\hbar \omega_c}{2k_BT})}{({1+n_{ph}/n_c})^\phi}\label{eq_sq},
\end{equation}
which depends on a loss constant $1/Q_i^0$, resonator frequency $\omega_c$, temperature $T$, and the crossover photon number $n_c=n_c(\vb*{p_i},  \vb*{E}_{\mathrm{zp}}(\vb*{r}),\tau_i)$, where the latter depends on TLS dipole moments $\vb*{p_i}$, the zero-point electric field $\vb*{E}_{\mathrm{zp}}$, and the TLS coherence time $\tau_i$.
Recent measurements of resonators with deposited films support the saturation power law with an exponent of $\phi=$0.5  \cite{martinis2005decoherence,lindstrom2009properties,pappas2011two,MoeaAl2O3,skacel2015probing}.
However, planar resonators with oxide interfaces and possibly also processing residues have a saturation exponent $\ll0.5$ \cite{khalil2010loss,macha2010losses,muller2019towards,Burnett_Res,Graaf_suppression2018}.

Generally, there are different possible causes for the weak power dependence ($\phi\ll0.5$). 
One possibility is that this is caused by strong TLS-TLS interactions which lead to fast TLS frequency switching. Instead, there may be wide distributions in: $p$ \cite{muller2019towards, hung2022probing}, $\vb*{E}_{\mathrm{zp}}(\vb*{r})$ \cite{khalil2010loss,wang2015surface,gorgichuk2022origin}, or $\tau$ \cite{palomaki2010multilevel}. Dielectric films sometimes exhibit a large distribution width in $p$ \cite{sarabi2016projected,hung2022probing,yu2021evidence} and planar resonators have a position-dependent field distribution of $\vb*{E}_{\mathrm{zp}}(\vb*{r})$ \cite{wang2015surface}, which results in an effective distribution of $n_c$. 

Recent experiments show substantial interactions in qubits, including strong TLS spectral diffusion and telegraphic noise \cite{klimov2018fluctuations}, and avoided-crossings \cite{lisenfeld2015observation}. These are beyond the standard tunneling model (STM) of TLSs, which should describe deposited films, and relates to early work in TLS-TLS interactions \cite{yu1988low,burin1998interactions}.
In the recent theory of Ref. \cite{InteractingTLS1}, the authors derive a fast jitter rate $\gamma$.
Their model describes how the energy of coherent TLS (cTLS) (in the regime of $\hbar\omega\gg k_B T$) depend on the states of neighboring low-frequency (LF) TLSs (with $\hbar\omega\leq k_B T$). The model uses LF TLSs with a 1/f spectral density to create the known 1/f noise spectrum of the cTLSs. As a result of the frequency jitter, the power dependence for TLS saturation is weaker than in the STM.
The stochastic fluctuations yield a logarithmic equation which appears phenomenologically as a low $\phi$ \cite{Burnett_Res,Graaf_suppression2018}. However, this data set was taken using planar resonators, where the above concerns from field and TLS distributions are difficult to distinguish (and separately analyze). For scientific understanding, it is important to verify the correct explanation by some means.

Novel experimental techniques have been recently developed to further understand TLSs.    
An experiment which injected a detuned pump tone to a resonator revealed the average TLS decoherence rate $\Gamma_2$ and the TLS coupling rate $g$ \cite{TT2020} (or Rabi frequency $\Omega$ \cite{TT2017}). 
Applying static strain \cite{StrainTransmission2015}, DC electric field \cite{hung2022probing,sarabi2016projected} or AC low frequency electric field \cite{MoeACEfield,burin2013universal,yu2021evidence} allows the extraction of $p$-distribution information.
Besides the conventional view of defects as atomic systems \cite{STM1,STM2}, an experiment of NbN resonators shows that trapped quasiparticles might be another possible origin of surface TLSs \cite{de2020two}.

Many studies of TLSs explore thermally grown aluminum oxide $\mathrm{AlO_x}$, which is used in most JJs and appears on the surface of Al. The loss tangent of $\mathrm{AlO_x}$ is found to be $\sim\ 2\times10^{-3}$ in large JJs and bulk dielectric \cite{martinis2005decoherence,deng2014characterization}. 
Recently, a qubit with a large JJ as a shunt capacitor suggests that transmons are not limited by junction loss.
 \cite{mamin2021merged}.
One set of transmon qubit studies finds that 40\% of the individual TLSs are located in the JJ barrier and the rest are on surfaces \cite{Lisenfield2021,Lisenfield2019}. In another study, the Al metal-air interface with its oxide is found to be the dominant loss mechanism in a qubit type \cite{Oliver2020Res}. There are very few studies of TLSs at the surface-air interface. 

Here we use vacuum-gap (VG) capacitors in resonators, where the metal-air (MA) interfaces (i.e., $\mathrm{AlO_x}$) within them are the dominant TLS-host materials.
Additionally, we fabricate the VG capacitors with a nearly uniform plate gap such that we probe with a nearly uniform ac electric field.
Our MA filling factor $F_{ma}$ is as large as 0.9\% assuming a 2 nm surface dielectric thickness and a dielectric constant of 10. This filling factor is ten times larger than in a typical planar resonator design \cite{Oliver2019Res}.
Our data show a weak $n_{ph}$-dependent loss and a weak $n_{ph}$-dependent Rabi frequency from a pumping tone, which allows us to rule out the possibility of multiple contributions, e.g., $p$, to these low slopes.
The power spectral density of resonator transmission rate $S_{21}$ at various temperatures indicates that the temperature-dependent TLS dephasing rate is affected by LF TLSs.
Lastly, the jitter rate $\gamma$ decreases after annealing the devices in vacuum at 300$^\circ$C but $Q_i^0$ is unchanged, indicating that the LF TLS density decreases but not the near-resonant TLS density.
The design of resonators also allows us to apply a DC electric field on the capacitors for the tuning of TLS energies.
From the data, we extract $p$ = $1.5^{+0.8}_{-0.6}$ D and $g$ on the order of 100 kHz.
However, individual TLSs are not observed during DC electric field biasing because of the large TLS jitter rate.

\section{Theoretical models of TLS loss}
According to STM \cite{STM1,STM2}, under the condition of a uniform ac electric field $E_0$, a single-value dipole moment $p_c$, the loss tangent from TLSs can be written as
\begin{equation}
    \tan\delta_{tls} = \frac{\pi P_0 p_c^2}{3\varepsilon\sqrt{1+E_0^2/E_c^2}} \label{eq_corrected_PD},
\end{equation}
where $P_0$ is the TLS spectral density, $\varepsilon$ is the dielectric constant and the critical field $E_c \propto 1/(p_c\tau)$.
LC resonators with parallel-plate capacitors are useful to study the loss tangent of dielectric films since there is \red{only one dominant TLS host volume $V$, giving $E_0 =  \sqrt{2{n_{ph}\hbar \omega_c}/{\varepsilon V}}$}.

\red{On the other hand, in a planar resonator design, $E_0$ is non-uniform and often the dominant interface contributing to material interface loss is unknown \cite{Oliver2020Res}.}
An integral over all contributions is necessary leading to
\begin{equation}
    \frac{1}{Q_i} =\frac{\intop_V\,\varepsilon(\mathbf{r})\, |E_0(\mathbf{r})|^2 \,\tan\delta_{tls}(\mathbf{r})\ d\mathbf{r}^3}{\intop_V\,\varepsilon(\mathbf{r})\, |E_0(\mathbf{r})|^2\ d\mathbf{r}^3},\label{eq_int}
\end{equation}
where $\mathbf{r}$ is the position  \cite{khalil2010loss,wang2015surface,gorgichuk2022origin}, and the spatial dependence to the field $E_0(\vb*{r})=2 E_{zp}(\vb*{r}) \sqrt{n_{ph}}$ is related to the zero-point fluctuation field $E_{ZP}$.
In a constituent volume $V_i$ [add integral sub $V_i d^3r$ formula], the TLS properties ($P_{0}$, $p$, $\vb*{E}_{\mathrm{zp}}$ and $\tau$) are assumed to be the same and we define the filling factor $F_{i} = \intop_{V_i}\,\varepsilon(\mathbf{r})\, |E_0(\mathbf{r})|^2\ d\mathbf{r}^3/\intop_V\,\varepsilon(\mathbf{r})\, |E_0(\mathbf{r})|^2\ d\mathbf{r}^3$.
Therefore, an approximate expression for analyzing different material volumes hosting TLSs is \cite{muller2019towards} 
\begin{equation}
    \frac{1}{Q_i} = \sum_i F_{i}\,\tan\delta_{tls,i} = \frac{1}{Q_i^0 \left(1 + n_{ph}/n_c\right)^\phi}.\label{eq_sum}
\end{equation}
After fitting many resonators to this formula, it is unclear if small $\phi$ is caused by a distribution in $\vb*{E}_{\mathrm{zp}}$ resulting from the resonator geometry, or a wide distribution in TLS parameters $p$ and $\tau$, or large intrinsic TLS jitter. The summation in the previous formula represents TLS populations with different contributions, and thus it may describe a multi-contribution (MC) model (see also the fits of $Q_i$ in Appendix \ref{appendix_Two_TLS}). When the jitter rate $\gamma$ of TLSs is larger than $\Omega$, the STM is no longer valid \cite{InteractingTLS1,burnett2016analysis,Graaf_suppression2018}.
Instead, theory gives a logarithmic function for the loss tangent 
\begin{equation}
    \tan\delta_{tls} =  \,P_\gamma\,\tan\delta^0\, \mathrm{tanh}\left(\frac{\hbar \omega_c}{2k_BT}\right)\, \mathrm{ln}\left(\frac{\gamma_{max}}{\Omega} + C_1\right).\label{log_eq}
\end{equation}
\red{Here, $\tan\delta^0$ is the intrinsic loss and $P_\gamma = \mathrm{ln}^{-1}(\gamma_{min}/\gamma_{max})$, where $\gamma_{max}$ and $\gamma_{min}$ are the maximum and minimum jitter rates of LF TLSs in the model (respectively), and $C_1$ accounts the $n_{ph}$-independent loss. Eq. \ref{log_eq} describes a fast switching (FS) model for unstable cTLS frequencies, and contrasts the MC model which is built upon the STM.}

The fits of MC and FS models, Eqs. \ref{eq_sum} and \ref{log_eq}, respectively, can be distinguished in this work.
Although Eq. \ref{log_eq} fits well in Ref. \cite{Burnett_Res,Graaf_suppression2018}, the authors are not able to isolate or distinguish loss from multiple interfaces, such that the MC interpretation may be correct. 
In this article, we probe resonators with a nano-gap VGC, and they are limited by TLSs in the surface oxide of Al and find that the FS model is appropriate to explain our various data types (including loss).

\section{Fabrication Method}

\begin{figure}[t]
\begin{centering}
\includegraphics[width=8.6cm]{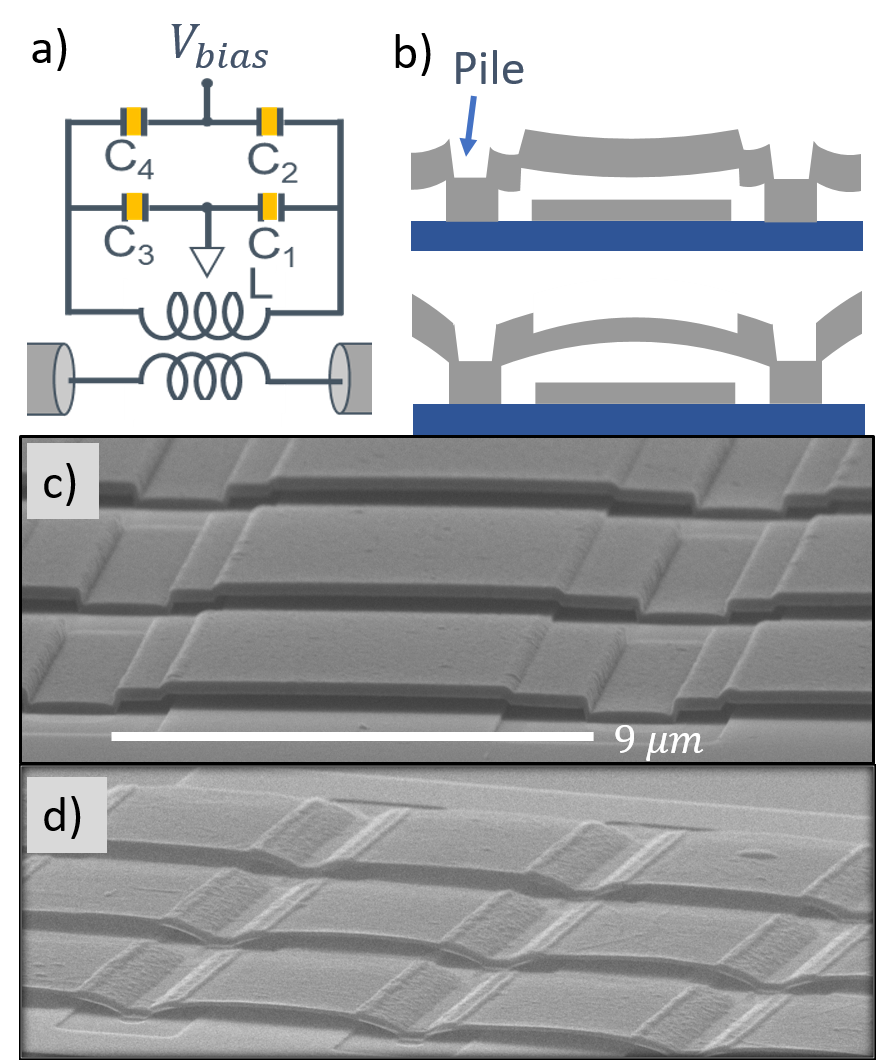} 
\par\end{centering}
    \caption{Schematic, side-views and scanning electron micrographs of vacuum-gap capacitors (VGCs). (a) Schematic of the resonators. (b) Side-view illustrations of the VGCs using SiNx (top) and photoresist (bottom) as sacrificial layers (SLs). The shape of the top electrode depends on the SLs (see main text). (c) VGCs with SiNx as the SL, which yields a terraced bridge shape with an approximate average gap distance $\bar{d}$ $\approx$ 125 nm. (d) VGCs with photoresist as the sacrificial layer, which yields an arch shape and $\bar{d} \approx$ 250 nm. The process uses a 200 nm metal-thinning step before the removal of the SL. Both types of VGCs have variation in $d$ of less than 15\%. \label{fig1_fab}}
\end{figure}

The schematic of the resonators follows Ref. \cite{sarabi2016projected,hung2022probing} and is shown in Fig. \ref{fig1_fab} (a). 
VG-based elements can provide a small on-chip footprint of capacitance and still maintain high quality for both micro- and mechanical wave resonators \cite{KatVG2010,JohnVG2011,palomaki2013entangling}. 
Vacuum-gap capacitors (VGCs) are achieved with a standard optical lithography process; two kinds of sacrificial layers (SLs) are used, which are low stress SiNx and photoresist (PR).
The capacitors are comprised of perpendicular strips with widths of 6 and 13 $\mu$m.
After releasing the SL, the top electrode strip shape ensures a quasi-uniform gap distance.
At the start of the fabrication, the bottom electrode and the ground plane are made from 120 nm-thick aluminum film deposited by e-beam evaporation to ensure a low stress film.  
Next, we deposit the SL and name the resonators \emph{SiNx VGC} and \emph{PR VGC}, depending on the use of the SL. 
For SiNx VGCs, we grow 200 nm-thick SiNx in an Oxford PECVD system at 300$^{\circ}$C and spin PR to pattern the vias and bridge supports.
Later, we etch the patterns by $\mathrm{SF_6}$ reactive-ion etching (RIE) plasma.
For PR VGCs, we spin S1805 positive PR at 500 nm thickness, and then expose the pattern for vias and bridges. Finally, we developed it with Microposit MF-CD-26.
We reduce the PR thickness from 500 to 200 nm by a timed RIE oxygen plasma descum and then bake it at 170$^{\circ}$C to strengthen the PR.
A top Al layer of 300 nm is deposited by e-beam evaporation and patterned to form the top electrode plate.
We choose 300 nm-thick Al in order to fill vias and strengthen the base piles for the top electrodes.
Side-view illustrations of two VGCs are shown in Fig. \ref{fig1_fab} (b).
An additional thinning process is made by etching 100 $\sim$ 200 nm of the top electrode to prevent the bridge from collapsing. 
The thinning process reduces the beam mass/thickness which changes the strain from compressive to tensile.
We show a SiNx VGC without the thinning process in Fig.\ref{fig1_fab} (c) and PR VGC with 200 nm thinning in Fig.\ref{fig1_fab} (d).
The thinning process is essential for PR VGCs generally and both types of VGC if the support piles have large separations. 
The separation of piles is chosen as 9 $\mu$m in the final devices, and a separation > 15 $\mu$m always collapsed after releasing.

Next, the wafers are diced into 6.5 X 6.5 mm$^2$ chips and the SL is ready to be released.
For PR VGC, the chips are immersed in 80 $^{\circ}$C NMP for 3 hours, rinsed with IPA, and blown dry by nitrogen. 
For SiNx VGC, the chips are placed under a high power inductively coupled $\mathrm{SF_6}$ plasma without forward voltage, which has a lateral etch rate of around 4 $\mu$m per hour.
Finally, we measure the resistivity from the bias port of the resonators and the ground to confirm a non-collapsed VGC, where the yield is more than 95\%.
The gap distance $d$ depends on the choice of the SL and the thinning process.
We expect the curvature or $d$ would change during cooling from room temperature to mK.
However, this effect would be small and in fact all non-collapsed VGCs did survive cooling to cryogenic temperatures.

The gap $d$ is estimated by comparing the measured resonant frequency to that of the finite-element microwave simulation in COMSOL.
We find that $d$ of all four resonators on the same chip have the same value.
However, because of the strain, the top electrode-bridges are never precisely parallel to the bottom electrodes (see Fig. \ref{fig1_fab} (b)).
By assuming $d$ is a quadratic equation, we obtain the coefficient of variation (CV) of $d$ is < 15\% for both types of VGC.
From the simulation, we found that the average gap distance $\bar{d}$ from SiNx VGCs is 125 nm for no thinning process (Fig. \ref{fig1_fab} (c)), 135 nm for a 100 nm thinning process, and 150 nm for a 200 nm thinning process. 
The $F_{ma}$ are (0.9\%, 0.8\%, 0.7\%), respectively, and $F_r$ of other interfaces is an order of magnitude lower than that in Ref. \cite{Oliver2019Res}.
While PR VGCs in Fig. \ref{fig1_fab} (d) have $\bar{d}$ = 225 nm and $F_{ma}$ < 0.45\%, which is smaller than our other resonators, it should have more contributions to interfaces other than MA relative to SiNx VGCs. 
Since the MA surface loss dominates, the surface loss tangent is $\tan\delta_{MA}=\left(Q_i\,F_{ma}\right)^{-1}$.

\section{TLS Loss and Noise Data}

\begin{figure*}[t]
\centering{}
\includegraphics[width=17cm]{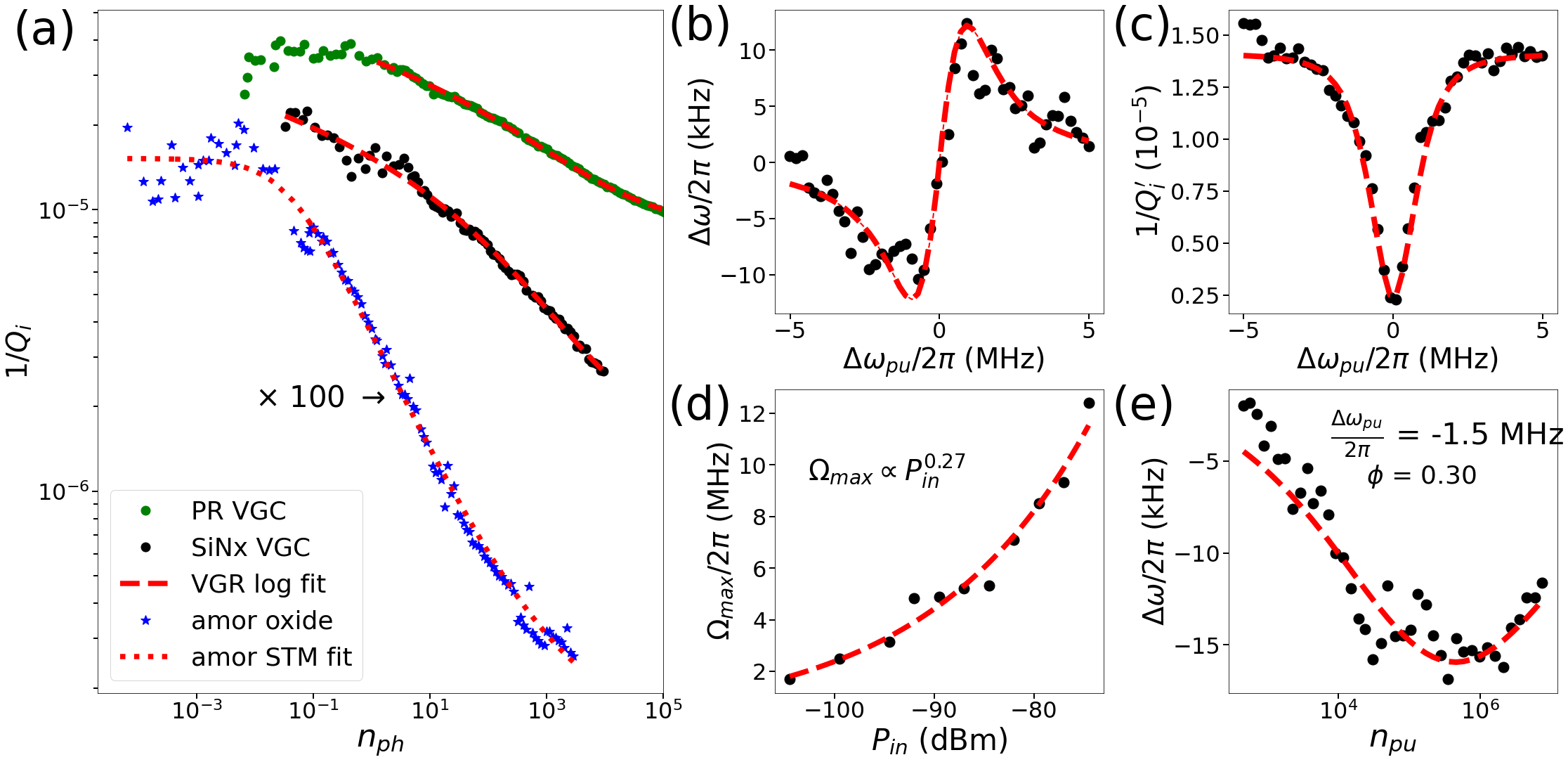}
\caption{
    (a) Intrinsic loss $Q_i^{-1}$ vs. average photon number $n_{ph}$. 
    The resonator results of a photoresist (PR) VGC, a SiNx VGC, and a bulk amorphous $\mathrm{AlO_x}$ sample are displayed from top to bottom, where the $\mathrm{AlO_x}$ loss is taken from Ref. \cite{hung2022probing} and is offset on the logarithmic scale such that multiplication by 100 restores the true values. 
    For the amorphous sample (blue), the single photon loss is 1.5$\times 10^{-3}$ and the saturation slope follows the standard TLS model (STM), where the fit to Eq. \ref{eq_corrected_PD} (shown) has the expected exponent of $\phi$ = 0.50.
    However, the VGC results fit to the fast switching (FS) model (Eq. \ref{log_eq}), which is a logarithmic function, and we have no need for a multiple contribution (MC) interpretation (see appendix for valid regime). 
    (b-e) Two-tone spectroscopy with a pumping tone frequency $\omega_{pu}$, probed around the original resonant frequency $\omega_c$.
    (b) Resonant frequency shift $\Delta\omega=\omega_c^\prime-\omega_c$ vs. pumping tone detuning $\Delta\omega_{pu}=\omega_{pu}-\omega_c$. 
    (c) $Q_i^{-1}$ vs. $\Delta\omega_{pu}$. 
    Panel (b) and (c) show the two-tone data from the SiNx VGC at a fixed input power $P_{in} \approx$ -105.5 dBm. 
    The red curves are the simultaneous least-squares fit of both $\Delta\omega$ and $Q_i$ to Eq. \ref{eq_freqshift} and Eq. \ref{eq_Qishift} except for a replacement of $n_{pu}^{0.5}\rightarrow n_{pu}^{\phi}$, where $\phi$ allows a deviation from the STM.
    From the fit, we obtain the maximum Rabi frequency of TLSs $\Omega_{max}=2\pi\cdot$1.7MHz. 
    (d) Plot of $\Omega_{max}$ vs. $P_{in}$, with fit line of $\Omega_{max}\propto P_{in}^{0.27}$. Since the exponent is < 0.5 and shows a weak input power dependence, the data are consistent with the FS model. 
    (e) $\Delta\omega$ vs. $n_{pu}$ for a fixed $\Delta\omega_{pu}=-2\pi\cdot$1.5 MHz (Eq. 7-9). The fit yields $\phi = 0.30$, which is consistent with the weak dependence described by the FS model. \label{fig2_PD}}
\end{figure*}

The microwave setup is given in Ref. \cite{hung2022probing}, and has a strongly attenuated input line and a low-pass filtered bias line.
Six chips are measured, consisting of 4 SiNx VGCs and 2 PR VGCs, and each chip contains four resonators.
All resonators worked with repeatable resonant frequencies between cooldowns, except one that may have a collapsed capacitor.
An internal quality factor in the single photon regime $Q_i^0\gtrapprox$ 60k is found in SiNx VGCs, regardless of the thinning process. 
$Q_i^0\gtrapprox$ 20k found in PR VGCs is possibly lower due to the ICP dry-etching thinning process causing extra residues.
A control group of coplanar waveguide resonators fabricated along with the VGC fabrication process shows a high internal quality factor, $Q_i^0$ > 0.5M, independent of the choice of SL (SiNx or PR), as expected. 
The SiNx VGCs have a MA surface loss tangent of $\tan\delta_{MA}=1.9\times 10^{-3}$, assuming the standard native oxide thickness of Al (2nm, as mentioned above). 
This is a loss tangent that is comparable to previous large-area loss measurements of AlOx mentioned above, and in a recent study of thin deposited films by the authors \cite{hung2022probing}. 
Thus we deem that the MA interface in SiNx VGCs is dominated by AlOx loss, and we focus on this VGC type in the analysis below.

Due to the unstable TLS frequencies, $Q_i$ fluctuates over time. 
Therefore, each $Q_i(n_{ph})$ in Fig. \ref{fig2_PD} is an average value from 10 acquisitions of transmission $S_{21}$ and the separation time of two acquisitions approximates 1.5 hours.
Weak $n_{ph}$ dependent losses of a SiNx VGC (black) and a PR VGC (green) are found in all VGCs, where $\phi$ is 0.18 $\sim$ 0.22.
Their traces fit well to the logarithmic Eq. \ref{log_eq} derived from the FS model as shown in the red dashed line.
$Q_i (n_{ph})$ of VGCs follows Eq. \ref{log_eq} up to a certain $n_{ph}$ but starts to increases afterward (see Appendix \ref{appendix_duff}). We only analyze data that is an order of magnitude smaller in photon number than where the loss reverses.
From the fit, we obtain $\gamma_{max}$ = $2\pi\,\cdot\,$5.7 MHz and $\gamma_{min}$ = $2\pi\,\cdot\,$20 kHz using $p$ = 1.5 (Debye) which is extracted later in Fig. \ref{fig2_PD} (d). 
We obtain $F_{ma} P_r \tan\delta^0\ln(C_1)<4\times10^{-7}$ indicating the $n_{ph}$-independent loss is trivial and not affected by the accuracy of the fitting.
In contrast, for the deposited $\mathrm{AlO_x}$ (obtained from our resonator with a capacitor filled with AlOx), we find a good fit to Eq. \ref{eq_sum} with $\phi = 0.5$, as shown in Fig. \ref{fig2_PD} (a) in blue dots.
This deposited-film $\mathrm{AlO_x}$ data came from a device used in a previous study, where we found some spectral diffusion and jitter from individual TLSs \cite{hung2022probing}.
However, since the power saturation of loss is strong and agrees with STM, it implies small mutual interactions between cTLSs and the LF TLSs, from the analysis of this measurement.
The larger $\gamma$ in MA TLSs compared to deposited-film TLSs is possibly due to a larger LF TLS density in its $\mathrm{AlO_x}$.

To further understand these MA TLSs, we perform two-tone spectroscopy, where the energy of the first tone is a weak probe ($\sim$ 0.1 $\hbar\omega_c$) and the second tone saturates TLSs at frequency $\omega_{pu}$ \cite{TT2017,TT2020}.
For a saturation tone detuned by $\Delta\omega_{pu}=\omega_{pu}-\omega_c$ > 0 from the original resonator frequency $\omega_c$, both the resulting resonance $\omega_c^\prime$ and internal quality factor $Q_i^\prime$ would increase because of the reduced TLS ground-state population on the upper side of $\omega_c$.
The complex frequency shift created by a single TLS is 
\begin{equation}
    \delta\omega + i \left(\frac{\delta\Gamma_i}{2}\right) =\frac{g^2}{(\omega_{tls}-\omega_c)+i\Gamma_2}\ \sigma_{z0}(\omega_{tls},\omega_{pu},\Omega),\label{complex}
\end{equation}
where the ground state population $\sigma_{z0}$ is a function of TLS frequency $\omega_{tls}$, $\omega_{pu}$ and $\Omega$ \cite{TT2020}.
The real and imaginary parts of Eq. \ref{complex} determine the shift in resonance $\delta\omega$ and internal decay rate $\delta \Gamma_i$, respectively, and the collective effect requires a summation over all TLSs.
The resulting frequency shift $\Delta\omega\ =\ \sum_{tls}\delta\omega = \omega_c^\prime-\omega_c$ and quality factor $Q_i^\prime$ \cite{TT2017} are written as
\begin{equation}
    \frac{\Delta\omega}{\omega_c}=\frac{3\pi}{4\sqrt{2}\,Q_i^0}\ X_\Delta\frac{\sqrt{1+X_\Delta^2}-1}{\sqrt{1+X_\Delta^2}+1} \label{eq_freqshift}
\end{equation}
and 
\begin{equation}
\resizebox{0.47\textwidth}{!}{$
    \frac{Q_i^0}{Q_i^\prime}=3X_\Delta^2\left[2+\sqrt{1+2X_\Delta^2}\ln\left(1+ X_\Delta^{-2} -\sqrt{1+2X_\Delta^{-2}}\right)\right]\label{eq_Qishift}$}.
\end{equation}
The dimensionless pump frequency $X_\Delta = {\Delta\omega_{pu}}/{\Omega_0}$ is dependent on the TLS Rabi frequency $\Omega_0$, which in turn is a function of photon number $n_{pu}$ in the cavity,
\begin{equation}
    \hbar\Omega_0=  {2\langle|\vb*{p}\cdot\vb*{E}_{\mathrm{zp}}|\rangle}\sqrt{n_{pu}},\label{eq_Omega_n}
\end{equation}
where the angle brackets $\langle...\rangle$ denote an expectation value.
The photon number resulting from the pump tone $n_{pu}$ can be obtained from the pumping power $P_{in}$ by 
\begin{equation}
    n_{pu} = \frac{2\kappa}{4\Delta\omega_{pu}^2 + \kappa_{tot}^2}\frac{P_{in}}{\hbar\omega_{pu}},\label{eq_num_photon}
\end{equation}
where $\kappa_{tot}$ and $\kappa = 2\pi\cdot 200$ kHz are the total and external decay rate of the resonator, respectively.
Typically, in planar resonator designs, the position dependence of $\vb*{E}_{\mathrm{zp}}$ obscures the extraction of $p$ from $\Omega_0$.
However, in our VGC most of the energy (>85\%) is in the gap such that $\vb*{E}_{\mathrm{zp}} \approx \sqrt{\hbar\omega_c/2Cd^2}\ \hat{z}$ \cite{girvin2014circuit}, where $\hat{z}$ is the normal direction to the plate.
Therefore, $p$ can be obtained from the two-tone technique.

From our results of the two-tone spectroscopy, we realize a similar phenomenological exponent $\phi$ is needed to fit the data ($n_{pu}^{0.5}\rightarrow n_{pu}^{\phi}$) and
\begin{equation}
    \hbar\Omega_0=2\langle|\vb*{p}\cdot\hat{z}|\rangle|\vb*{E}_{\mathrm{zp}}|\, n_{pu}^\phi\label{eq_rabi_correct}.
\end{equation}
First, we show $\Delta\omega$ vs. $\Delta\omega_{pu}$ (Fig. \ref{fig2_PD}b) and $1/Q_i^{\prime}$ vs. $\Delta\omega_{pu}$ (Fig. \ref{fig2_PD}c) at a fixed $P_{in}\approx$ -105.5 dBm.
The red curves are the simultaneous least-squares fit of both $\Delta\omega$ and $Q_i^\prime$ to Eq. \ref{eq_freqshift} and Eq. \ref{eq_Qishift}. 
From the fit, we obtain $\phi$ = 0.3 and $\Omega_{max}=\Omega_0(\Delta\omega_{pu}=0)=2\pi\cdot$1.7MHz, which is also the maximum of Rabi frequency at $\Delta\omega_{pu}=0$ and at a fixed $P_{in}$.
We plot $\Omega_{max}$ vs. $P_{in}$ in Fig. \ref{fig2_PD} (d), which leads to $\phi$ = 0.27 and is consistent with a FS model ($\phi$ < 0.5).
Similarly, we measure the $1/Q_i^{\prime}$ and $\Delta\omega$ vs $n_{pu}$ at a fixed $\Delta\omega_{pu}=2\pi\cdot$1.5 MHz, and the fit yields $\phi = 0.3$ in Fig. \ref{fig2_PD} (e). 
It is worth mentioning that $\kappa_{tot}$ is dependent on $Q_i^\prime$, which in-turn depends on $\Delta\omega_{pu}$. 

Similar to Eq. \ref{eq_sq}, which phenomenologically describes slow saturation from a pump with $\phi$, we also substitute $n_{ph}^{0.5} \rightarrow n_{ph}^{\phi}$ with $\phi$ < 0.5 for all equations derived from the STM including Eqs. \ref{eq_sum}, \ref{eq_freqshift}, \ref{eq_Qishift}, and \ref{eq_rabi_correct}.

This is a second use of the low $\phi$ besides the MC model (which assumes a distribution of parameters, such as a distribution in $p$). However, we found that the MC model (see Appendix \ref{appendix_Two_TLS}) yielded a worse fit than Eq. \ref{eq_Omega_n} so it appears that only one group of $p$ exists and it exhibits low power saturation. Assuming that the TLS distribution d$n\propto 1/\tau$, which is dependent on the coherence time $\tau$ (\cite{palomaki2010multilevel}) for a single value of $p$, we find $\phi = 0.5$ (similar to \cite{InteractingTLS1}). As a result, we rule out the MC model and find that the FS model is qualitatively consistent.

From Eq. \ref{eq_num_photon} and \ref{eq_rabi_correct}, we extract $p$ = $1.5^{+0.8}_{-0.6}$ Debye = $0.3^{+0.18}_{-0.1}$ e$\mathring{A}$, where consider isotropic angle distribution $\vb*{p}$, $\langle|\vb*{p}\cdot\hat{z}|\rangle = p/\sqrt{3}$ \cite{Gaothesis} and that the precision is limited by gap non-uniformity. The gap non-uniformity and accuracy is discussed in more detail in Appendix. \ref{COMSOL}. For comparison, our two-tone spectroscopy is consistent with Ref. \cite{Ustinov2021Acous}, where they resolved $\phi$ = 0.28 $\sim$ 0.3 in a surface acoustic wave resonator with the same method as we used to collect data in Fig. \ref{fig2_PD} (d).

\begin{figure}
\centering{}
\includegraphics[width=8.6cm]{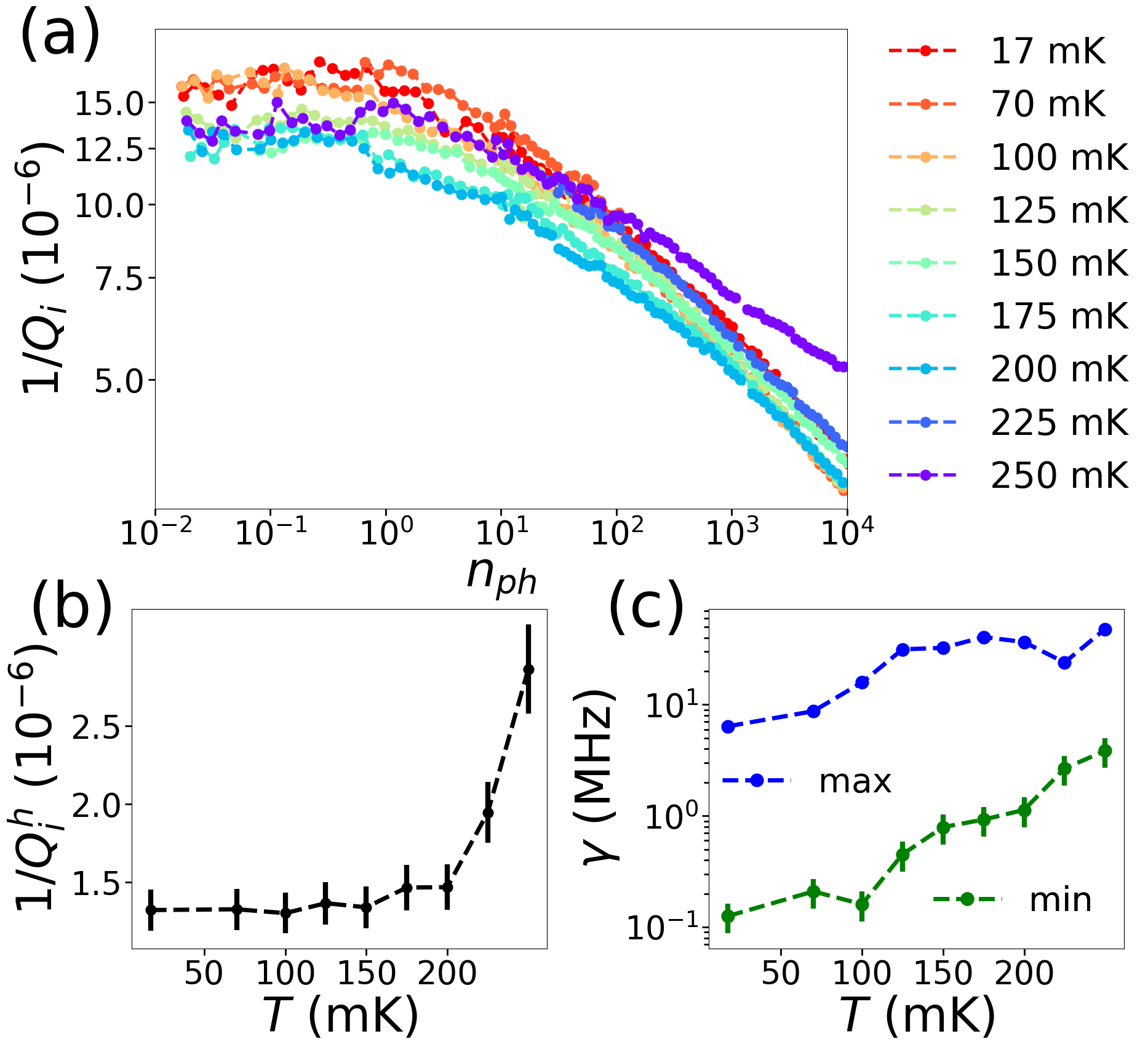}
\caption{
    (a) $1/Q_i$  vs. $n_{ph}$ in a SiNx VGC resonator at different temperatures. The exponent of $Q_i^{-1}(n_{ph})$ ($\phi$) slightly decreases while temperature increases. (b) The high power background loss is defined as $1/Q_{i}^h = 1/Q_i(n_{ph} = 10^4)$ for each temperature (see text). This quantity is relatively constant for low temperature < 200 mK and increases at higher temperatures due to larger numbers of quasiparticles. (c) The extracted maximum and minimum jitter rate $\gamma_{max,min}$ increases with temperature. This is expected since $\gamma$ is affected by the number of thermally activated low-frequency (LF) TLSs. \label{fig3_temperature}}
\end{figure}

\red{We show the temperature $T$ dependence of the resonator loss $1/Q_i$ in Fig. \ref{fig3_temperature} (a). The loss is predominantly from MA interface TLSs, and at the lowest photon number the TLS loss is largest for the lowest temperatures. However, there is also loss seen at high photon number $n_{ph}$ = $10^4$ and T = 250 mK, where the photon number saturates the TLSs and the background loss remains, $1/Q_i^h = 1/Q_{i,0} - 1/Q_i(n_{ph} = 10^4)$. Some of this loss is from temperature-induced quasiparticles in the aluminum wiring. To account for the high power loss, we fit this data using Eq. \ref{log_eq}, and from $C_1$ found that $1/Q_i^h$ versus temperature in Fig. \ref{fig3_temperature} (b). It reveals large quasiparticle loss in two data points at $T\ \geq$ 200 mK. The fit also gives the maximum and minimum jitter rates $\gamma_{max}$ and $\gamma_{min}$ of the LF TLS distribution, respectively, as shown in Fig. \ref{fig3_temperature} (c). The data show an increasing maximum jitter rate with increasing temperature and a minimum rate that is always larger than the Rabi frequency.}

\begin{figure}
    \includegraphics[width=8.6cm]{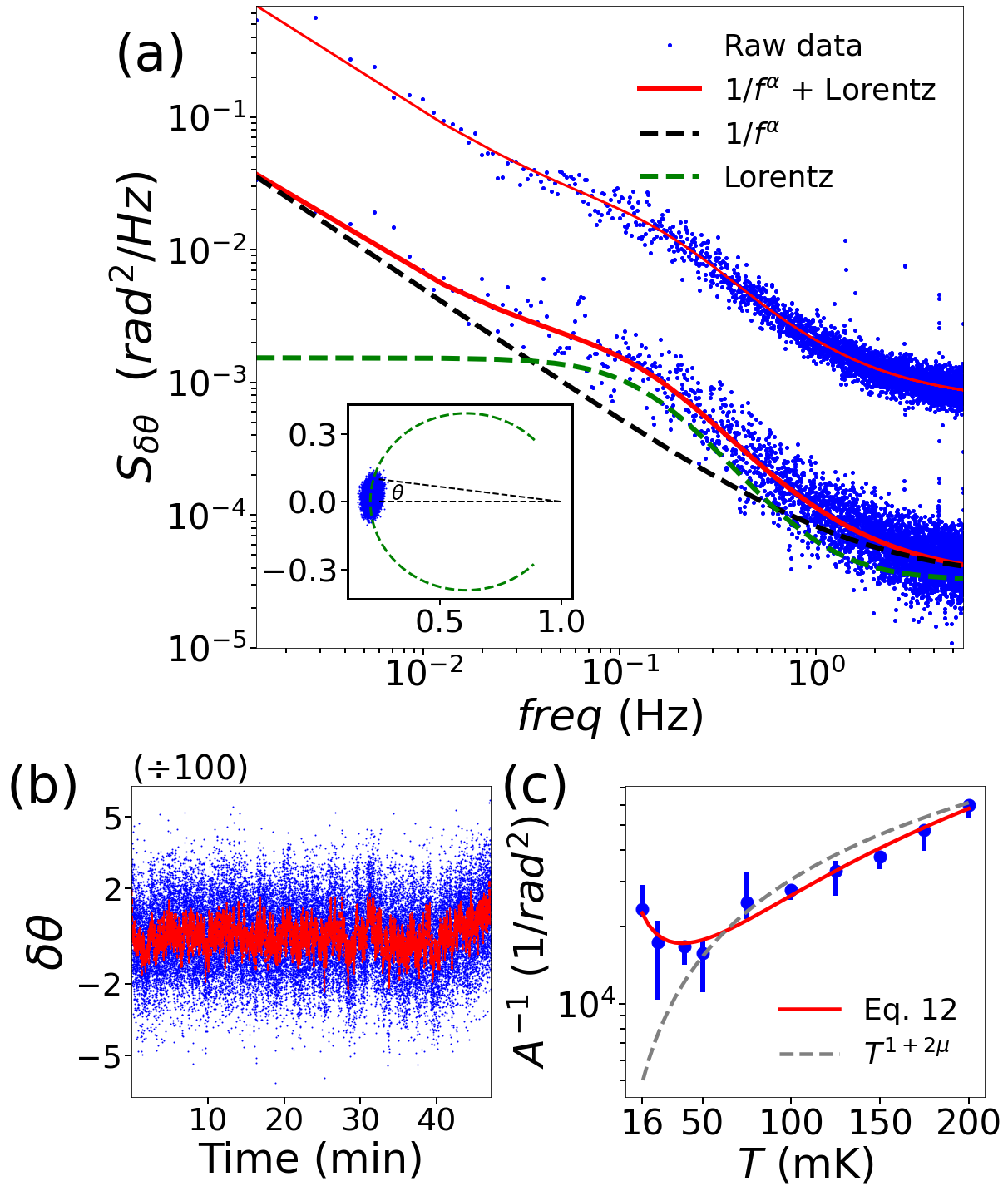}
    \caption{Noise measurements from the SiNx VG resonator: (a) Power spectral density (PSD) of phase angle $\delta\theta = \theta - \bar{\theta}$, where $\theta$ is centered near zero in noise measurements: The inset shows the transmission data at one frequency on an I-Q plot of the resonance, where $\bar{\theta}$ is the mean value of $\theta$ in a dataset. PSD is fit to a 1/f spectrum with Lorentz function and white noise ($A/f^\alpha\ +\ B \tau/(1+\left(2\pi f\tau\right)^2)+C$) in the red curves. An example trace and an average of 8 separate measurements (shifted by 50 times) are shown on the bottom and top, respectively. (b) A plot of $\delta\theta$ vs time: The blue dots are the raw data, and we applied a Gaussian filter in time to obtain the set of smoothed red dots. (c) The inverse of $A$ from the averaged PSD vs. temperature $T$: The error bar represents the max and min of $A$ in 8 separate measurements at each $T$.  We assume the TLS decoherence rate is temperature dependent, $\Gamma_2 \left( T \right)= \Gamma_{1}/2+\Gamma_{\phi}\left(T\right)$, where the LF-TLS induced dephasing rate $\Gamma_{\phi}\left(T\right)\propto T^{1+\mu}$. From the red fit to Eq. \ref{eq:noise}, we obtain $ \mu $ = 0.3 $\pm$ 0.2, which is consistent with the FS model (see main text).}
    \label{fig4_noise}
\end{figure}

According to Ref. \cite{InteractingTLS2}, cTLSs' decoherence rate $\Gamma_2\propto T^{1+\mu}$ depends on the number of coupled LF TLSs.
Here, $\mu$ describes a non-uniform TLS asymmetry energy $\Delta$ density: $P(\Delta)\propto\Delta^\mu$.
This density of states dependence is introduced phenomenologically and is stronger than the expected dipolar gap from TLS-TLS dipolar interactions \cite{burin1998interactions}.
Additionally, previous measurements \cite{ramanayaka2015evidence,Graaf_suppression2018} of 1/f resonant frequency noise induced by cTLSs show that the noise $\propto\,\Gamma_2^2/T = T^{1 + 2\mu}$ with $\mu$ $\sim$ 0.3.
To study $\Gamma_2$ of MA TLSs, we perform a transmission $S_{21}$ measurement on resonance at various temperatures and record the power spectral density  (PSD) of the phase $\theta$. 
$\theta$ is defined as the angle between a vector defined from the off-resonant transmission (defined as (1,0)) to the transmission ($S_{21}(t)$) and the x-axis (see inset of Fig. \ref{fig4_noise} (a)).
The PSD of the fluctuations in the phase angle, $\delta\theta$, provides the information of the fluctuations in the resonator frequency \cite{Gaothesis}.  
The lower trace of the panel (a) shows one data set of the PSD of $\delta\theta$. 
We fit this PSD to a function combining 1/f, Lorentz and white noise: $S_{\delta\theta} = A/f^\alpha\ +\ B \tau/(1+\left(2\pi f\tau
\right)^2)+C$. 
The Lorentz noise indicates a single or a few coupled cTLSs with $\tau\ \approx$ 3 sec. 
\red{The low-frequency telegraph-type switching of these TLSs is in the band of our resonator measurements, and typical qubit measurements find higher frequency Lorentzian noise due to a higher frequency band in those measurements \cite{benchmarkingT1drop,TLSnoiseonqubit2}.}

The remaining noise is considered 1/f type because the fit yielded $\alpha = 0.9 - 1.1$; this arises from many weakly coupled cTLSs \cite{benchmarkingT1drop,niepce2021stability}.
Guidelines show the Lorentz noise and 1/f noise separately in dashed lines. 
In Fig. \ref{fig4_noise} (b) we show the same data in the time domain ($\delta\theta$ vs. time) (blue) and the same data after a Gaussian filter in the time domain (red). 
These data are consistent with the panel-(a) data analysis, and exhibit clear spectral diffusion and possible telegraphic noise. 
The noise spectrum is also fluctuating with time (and the same phenomenon in $Q_i$). In some data, the Lorentz noise switches on and off in time, and wherein the latter case it is covered by 1/f noise. To show this we average 8 consecutive measurements and display the result in the upper trace in Fig. \ref{fig4_noise} (a) with a vertical shift ($\times50$ on a log scale).
From this data, the average 1/f noise amplitude ($A$) from the eight scans is obtained and plotted in Fig. \ref{fig4_noise} (c), along with the minimum and maximum amplitudes which are shown by vertical bars.

Fig. \ref{fig4_noise} (c) shows the measured 1/f noise amplitude, $A$. Current TLS noise theory gives $1/A\propto {\Gamma_2^2}/{T}$ and $\Gamma_{2} (T)\propto T^{1+\mu}$. From this formula for $A$ we obtain A = 1.7  $\pm$ 0.2 $\times 10^{-5}$ using our data at $T\ >\ 80$ mK shown (see the gray dashed line in the figure). This magnitude is equivalent to A= 1.4  $\pm$ 0.2 $\times 10^{-14}$ for the proportional quantity defined another group: $S_{\delta y
} = S_{\delta \theta} / 16 Q^2$ \cite{Graaf_suppression2018}. 
\red{Our value is within a factor of two of 2.4 $\times 10^{-14}$, the value found in Ref. \cite{Graaf_suppression2018}, and a few times larger than 3.5 $\times 10^{-15}$, as found in Ref. \cite{burnett2018noise}. }

Our fit value of $\mu$ is 0, however, the goodness of the fit to the standard function is poor. 
\red{On the other hand, we find a better fit of $1/A$ to a similar phenomenological function,
\begin{equation}
    \frac{1}{A} =  \frac{(M + N\,\cdot\,T^{1+\mu})^2}{T},\label{eq:noise}
\end{equation}
where the numerator is proportional to $\Gamma_2^2$, and $M$ and $N$ are constants. 
This equation can be understood by considering that $\Gamma_2(T)$ has a $T$-independent term $\Gamma_{2,0}$ which gives  $\Gamma_2 (T)= \Gamma_{2,0} + \Gamma_{\phi} (T)$, and a mutual-TLS-induced dephasing rate remains as $\Gamma_{\phi} (T)\propto T^{1+\mu}$.
The fit is shown as a red line and yields $\mu$ = 0.375 $\pm$ 0.2.
Related to Eq. \ref{eq:noise}, Ref. \cite{lucas2022quantum}finds insufficient thermalization at the lowest temperatures and a flat background for $A$ when $T$ is below 80 mK.
Another possible explanation is that the relaxation rate, $\Gamma_1$, of our TLSs are higher than others such that $\Gamma_{2,0}$ = $\Gamma_1/2$ is not ignorable. 
Since both explanations are possible in our devices, an additional measurement of surface-TLS coherence using a qubit may be necessary to ultimately understand this.}

\begin{figure}
\includegraphics[width=8.6cm]{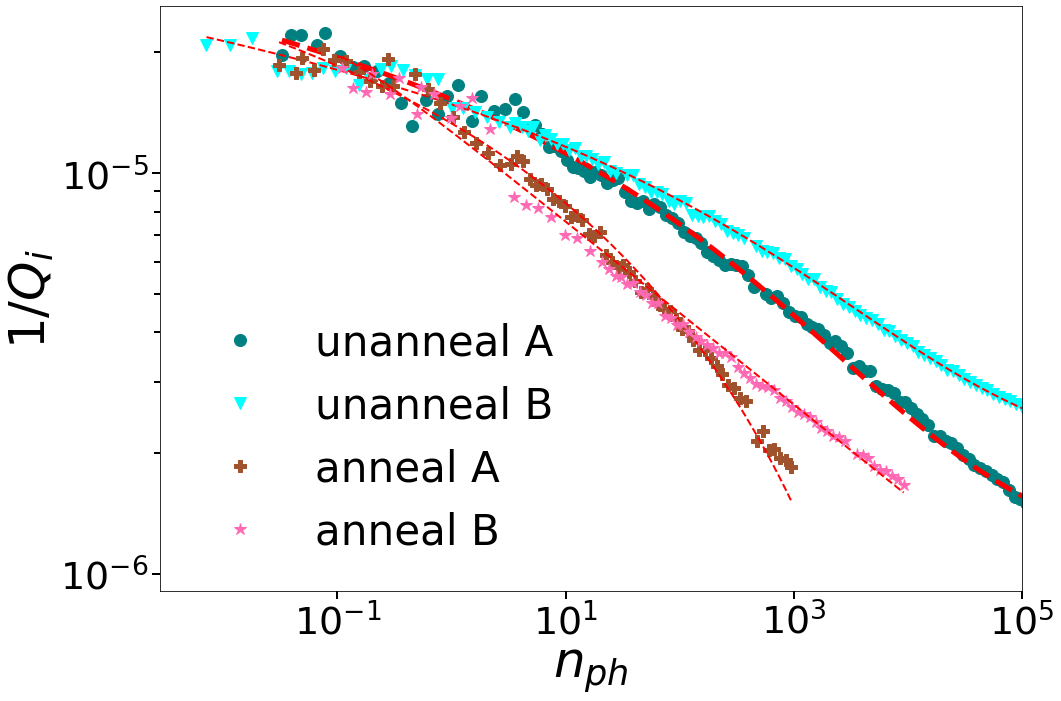}
\caption{$1/Q_i$ vs. $n_{ph}$ of unannealed and annealed SiNx VGCs. The red dashed lines are the fits to the Eq. \ref{log_eq}. We observe that $\gamma_{max}$ of the annealed VGCs are smaller than the unannealed VGCs. However, there is no improvement in single photon $Q_i^0$.
The probable cause to the decrease in $\gamma_{max}$ and unchanged $Q_i^0$ is that the desorption of surface spins belong to the set of LF TLSs, but not coherent TLSs.
\label{fig5_anneal}}
\end{figure}

We also performed additional measurements of four measured SiNx VGCs resonators after an annealing process was added. This post-annealing was performed under high vacuum at 300$^\circ$C for 1 hour duration. After annealing, three of the four resonators were visible in spectroscopy, and the fourth likely had a collapsed capacitor. 
The gap distance is $d\approx$ 125 nm and approximately constant in the majority of the devices, including the ones we report on in detail. We plot the 2 samples that showed the smallest and largest change in distance. The post-anneal distance is $d^\prime\approx$ 125 nm in Sample A, which showed no change in distance and Sample B showed a change to 100 nm in distance. The quality factors of these resonators, before and after the anneal are shown in Fig. \ref{fig5_anneal}.
Prior to the anneal, the resonators show an approximately constant loss in different cooldowns as they age (see Appendix \ref{appendix_aging}). 
We show the results of two SiNx VGCs, prior to and after annealing, and their fits to the FS model in Fig. \ref{fig5_anneal}. 
We find that the $Q_i^0$ and $A$ are not changed significantly by annealing, in contrast to Ref. \cite{Graaf_suppression2018}.
However, one obvious change is the slope of $Q_i^0$ versus $n_{ph}$.
For post-anneal A, $\gamma_{max}$ and $\gamma_{min}$ are changed from $2\pi\times$(5.7 MHz, 20 kHz) to  $2\pi\times$(1.9 MHz, 25 kHz) and, the value for post-anneal B is changed from $2\pi\times$(2.5 MHz, 3.3 kHz) to $2\pi\times$(0.9 MHz, 11 kHz). 
After annealing, we find that the number of photons needed for $Q_i$ is reduced by two orders of magnitude. 
\red{The improvement is likely caused by a reduction in the effective density of LF TLS due to the desorption of surface spins or rearrangement of surface atoms.
It is also possible that the LF TLS density distribution as a function of $\gamma$ is altered but not captured in the theory \cite{privatecom}.} 

\section{Discussion}
The dimensionless parameter describing collective mutual TLS-TLS interaction, $\chi = P_{0,LF} U_0\approx 10^{-2}\sim 10^{-3}$, is generally small in bulk materials and deposited films, which is critical for the steady-frequency approximation in the STM \cite{burin1998interactions}.
Here, $P_{0,LF}$ is the LF TLS density and $U_0$ is the characteristic interaction factor of a (typical) TLS pair.
In accordance with this, $Q_i$ in our resonators containing deposited amorphous $\mathrm{AlO_x}$ films is proportional to $\sqrt{n_{ph}}$ such that we expect that $\chi$ is small regardless of the observed TLS spectral diffusion in deposited films \cite{hung2022probing}.
However, this study finds that the MA-TLS phenomena agree with the FS model.
Thus, we expect that $\chi$ is larger on the Al surface than in deposited $\mathrm{AlO_x}$ films.
The extracted $p$ from the Al surface is $1.5^{+0.8}_{-0.6}$ D, which is only a few times smaller than $p$ in a typical deposited amorphous film.
In the standard theory, $U_0$ is expected to be similar in both samples if $U_0$ depends only on electric dipole moment and the dielectric is thick enough that the typical bulk theory applies.
Furthermore, from the post-annealing experiment, we learn that $P_{0,LF}$ is not necessarily equal to the density of cTLSs $P_{0,c}$, since $\tan\delta^0 \propto P_{0,c}$ is unchanged and $\gamma_{max}$ is lowered via the annealing.
Thus, two possible explanations for a larger $\chi$ of surface TLSs are that 1) the surface modifies the theory and 2) there is a much larger $P_{0,LF}$ at the Al surface than in a deposited $\mathrm{AlO_x}$ film.

Ref. \cite{mcrae2020materials} reviews the interface losses from a wide variety of materials, substrates and processes through measurements of either lumped or coplanar waveguide resonators. 
The collective effect from different interfaces brings complexity and it is generally impossible to distinguish the effects from different interfaces.
In the study of Refs. \cite{Oliver2019Res,Oliver2020Res}, the loss was extracted from different interfaces using different samples, but the power dependence of the loss of each interface was not extracted. 
Similar problems are found in the experiments of planar resonator frequency noise made from various superconductors \cite{barends2009noise}.

\section{Conclusion}
In the last decade, coplanar resonator studies have been important to the development of quantum information science. 
However, the loss of coplanar resonators limited by TLSs has a weak photon number dependence $1/Q_i \propto n_{ph}^{-\phi}$, where $\phi$ is a phenomenological fitting parameter.
\red{$\phi$ is measured at 0.18 - 0.22. 
This value is much smaller than the value, 0.5, derived from the standard TLS model (STM), which describes near-resonant TLSs as approximately steady in frequency.}
Coplanar resonators have multiple imperfect interfaces, which can host quantum defects.
The STM can possibly explain this phenomenon as a multi-contribution (MC) model if multi-interfaces have different TLS types (dipole moment, relaxation rate) or possibly if a broad distribution of electric fields is included from the resonator geometry.
In contrast, a fast-switching (FS) TLS model, considering mutual interacting TLSs, yields a logarithmic power dependence (Eq. \ref{log_eq}).
Previous studies could not distinguish between these two models.

To understand the loss without multiple interface types or distributed fields, we studied resonators with VGCs. 
These allow us to probe MA-interface TLSs with an approximately uniform ac field due to approximately parallel metallic plates.
We find that VGCs fabricated with an SiNx sacrificial layer have an internal quality factor of $Q_i^0\approxeq$ 20k which is consistent with AlOx loss. 
The weak $n_{ph}$ dependent loss is best fit to a FS (fast switching) model with one TLS type.
This contrasts the loss properties of $\mathrm{AlO_x}$ films, which agree with the STM. 
In conclusion, using VGCs we find that surface-based TLSs are distinguishable from bulk TLSs.

Next, we apply a two-tone spectroscopy and find the maximum TLS Rabi frequency $\Omega_{max}$ is weakly dependent on the input power $P_{in}$. 
All of the equations used to analyze data are adjusted from the STM, such that we generalize the power dependence as $n_{pu}^{0.5}\rightarrow n_{pu}^{\phi}$, including Eq. \ref{eq_corrected_PD}, \ref{eq_freqshift}, \ref{eq_Qishift}, and \ref{eq_rabi_correct}. The last 3 equations were needed to analyze two-tone spectroscopy data, and all give $\phi\ll0.5$.
Therefore, we can exclude the MC model which predicts $\phi$ = 0.5. Based on our data from a quasi-parallel plate VGC, which gives a weak $n_{ph}$ dependence to $Q_i$ and $\Omega_{max}$, we conclude the TLSs are switching frequencies at a high rate and that the TLS dipoles have a $p_z$-distribution width much smaller than its mean value.

We observe that the jitter rate of TLSs increases with temperature $T$ from the measurement of loss tangent.
Additionally, we observe an increase of TLS dephasing rate upon raising $T$ from the resonant phase noise. 
These effects are qualitatively consistent with a model where noise from LF TLSs affect high-frequency (near resonant) TLSs \cite{InteractingTLS2}.
Furthermore, we reduce the density of LF TLSs by post-annealing in vacuum at 300$^{\circ}$C.
As expected, this did not change the internal quality factor, which implies the source of low- and high-frequency TLSs are different.

Our study finds that surface TLSs, hosted in the MA (Metal-Air) interface on Al, are substantially different than TLSs hosted in deposited films. Specifically, TLSs in our vacuum gap capacitors, in the analysis of power-dependent loss, two-tone saturation, and the frequency noise spectrum, exhibit more jitter than a standard deposited AlOx. This work is a significant step towards isolating and studying individual TLSs on a metal surface in a uniform ac electric field.

\section{Acknowledgements}

We thank K. Cicak and R. Simmonds for scientific discussions on vacuum-gap capacitors. 
\renewcommand\thefigure{\thesection.\arabic{figure}}    
\setcounter{figure}{0}   
\setcounter{section}{0}
\appendix

\section{Alternative logarithmic equation and multi-contribution model}
\label{appendix_Two_TLS}
\begin{figure}
    \centering
    \includegraphics[width=8.6cm]{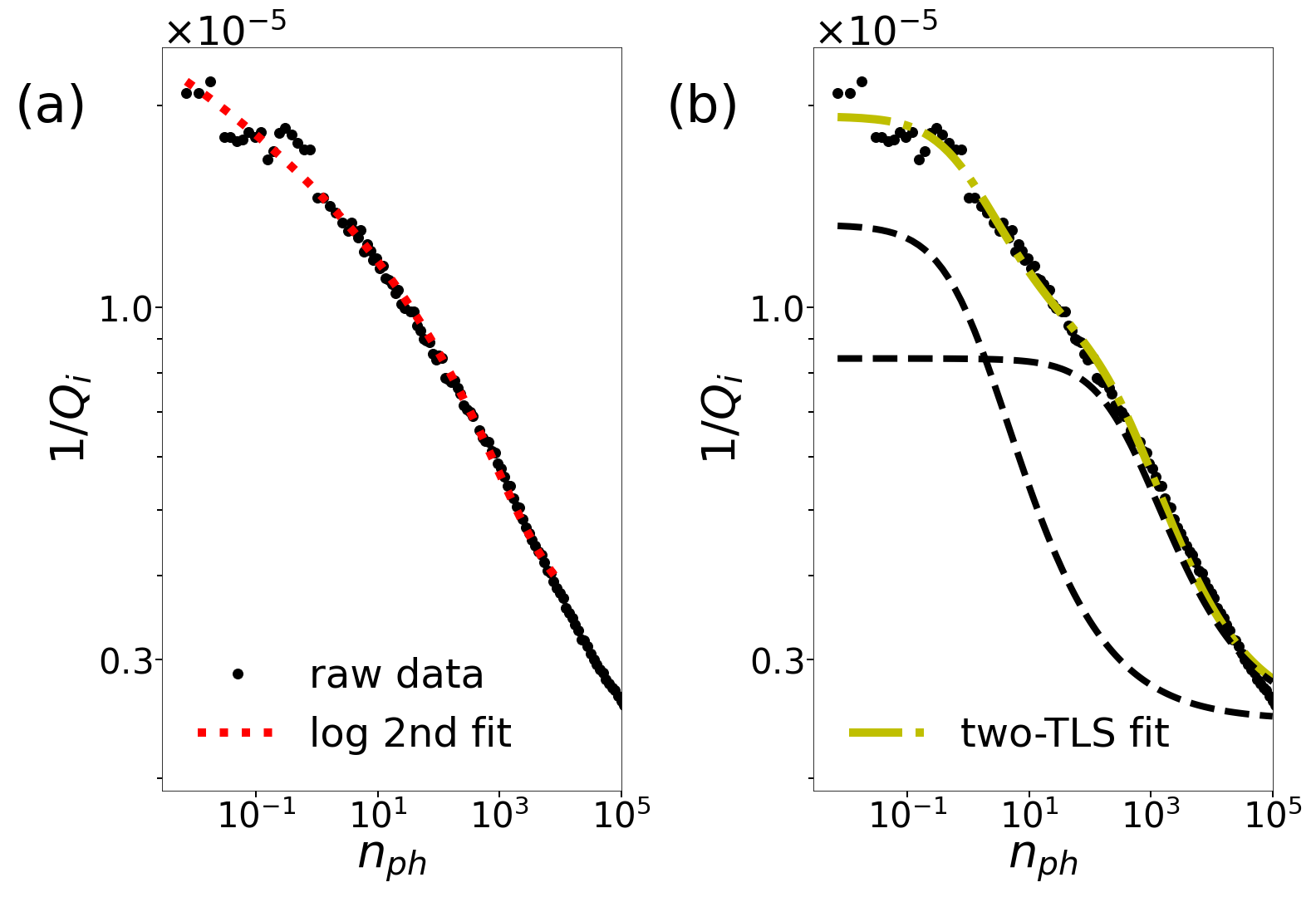}
    \caption{$1/Q_i$ vs. $n_{ph}$ for two different fitting equations. (a) Fit to Eq. \ref{log_eq2} using $p$ = $1.5^{+0.8}_{-0.6}$ Debye as given in Fig. \ref{fig2_PD} (d). The fit yields the geometric average $\sqrt{\gamma_{max}\Gamma_2}$ = $2\pi\cdot$ 14MHz. (b) Fit to Eq. \ref{eq_two_tls}. We obtain the fitting parameters of $F_{1}\tan{\delta_1^0={1.1\times10}^{-5}}$, $F_{2}\tan{\delta_2^0={6.0\times10}^{-6}}$, $n_{c,1}$ = 0.9, and $n_{c,2}$ = 320. The two dashed guidelines represent the contributions from each TLS contribution. However, from Fig. \ref{fig2_PD} (d), we exclude the possibility that Eq. \ref{eq_two_tls} is the correct equation (see the main text). }
    \label{sfig_extra_fit}
\end{figure}
\begin{figure}
    \centering
    \includegraphics[width = 5.5cm]{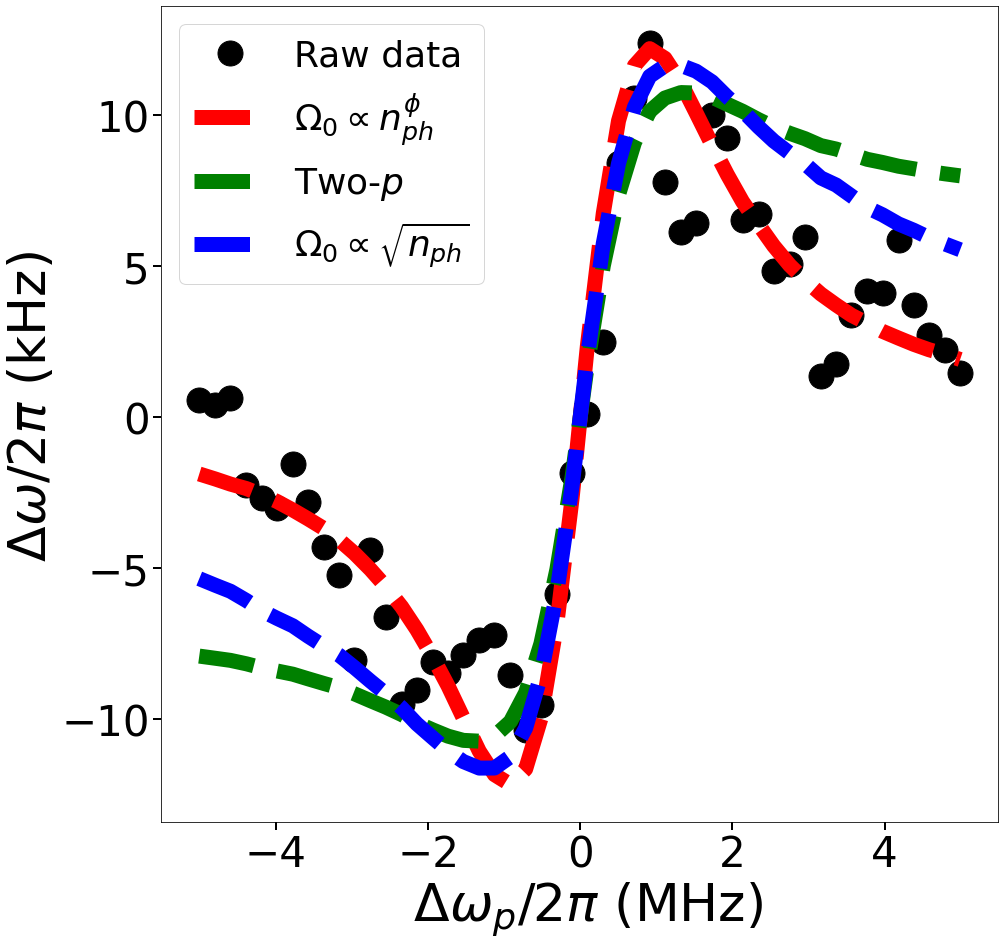}
    \caption{Two tone spectroscopy, $\Delta\omega$ vs. $\Delta\omega_{pu}$, using different models to fit. The black dots and the red curve are the same as in Fig. \ref{fig2_PD} (b). The blue and green curves are based on STM with one and two groups of TLSs, respectively. We find that only a weak power dependence $\Omega_0$ can explain the behavior of $\Delta\omega$ while increasing $\Delta\omega_{pu}$.}
    \label{sfig:two_TLS_TT}
\end{figure}
In the main text we chose to fit using Eq. \ref{log_eq}, which has the condition, $\gamma \gtrsim \Omega \gg \Gamma_1,\ \Gamma_2$ \cite{InteractingTLS1}. In this appendix, we first discuss two other possible models for fitting. The first equation derived in Ref. \cite{InteractingTLS1} is 
\begin{equation}
    \mathrm{tan}\delta_{i} =  \mathrm{tan}\delta_{i}^{0}\, P_\gamma \, \mathrm{tanh}(\frac{\hbar \omega_c}{2k_BT})\, \mathrm{ln}(\frac{\gamma_{max}\Gamma_2}{\Omega^2} + C_1),\label{log_eq2}
\end{equation}
and is in a particular regime defined by $\gamma\cdot\Gamma_2 > \Omega^2$. A fit to Eq. \ref{log_eq2} is shown in Fig. \ref{sfig_extra_fit} (a), and we extract the geometric average $\sqrt{\gamma_{max} \Gamma_2}=2\pi\cdot$14 MHz, which is much larger than $\Omega$ of the fitting range.
However, since we do not know the value of $\Gamma_2$, we prefer Eq. \ref{log_eq} in the main text.

On the other hand, the MC model is given by Eq. \ref{eq_sum} with $i$ = 1 to N, while a more accurate integral form is shown in Eq. \ref{eq_int} and can be also found in Ref. \cite{MoeaAl2O3,gorgichuk2022origin}.
Here, we consider the case of two contributions ($i$=1, 2), giving
\begin{equation}
    \mathrm{tan}\delta (n_{ph}) = \frac{F_{1}\ \mathrm{tan}\delta_{1}^0}{\sqrt{1+\frac{n_{ph}}{n_{c,1}}}} +\frac{F_{2}\ \mathrm{tan}\delta_{2}^0}{\sqrt{1+\frac{n_{ph}}{n_{c,2}}}} + C_0. \label{eq_two_tls}
\end{equation}
From the fit in Fig. \ref{sfig_extra_fit} (b), we obtain $F_{1}\tan{\delta_{1}^0={1.1\times10}^{-5}}$, $F_{2}\tan{\delta_{2}^0={6.0\times10}^{-6}}$, $n_{c,1}$ = 0.9, and $n_{c,2}$ = 320.
Two guidelines showed two separate square root dependent $Q_i$. 
According to the STM \cite{STM1,STM2}, $n_{c,i}^{-1} \propto (p_i {E}_{\mathrm{zp}})^{2}\tau$, $p_i$ is the dipole moment of i-th group. 
The fit yields $n_{c,2}/{n_{c,1}}\approx$ 320, which is valid only if there are two types of TLSs. 

However, the validity is negated by the results of the two-tone spectroscopy measurement.
In Fig. \ref{sfig:two_TLS_TT}, we observe that Eq. \ref{eq_rabi_correct} (red) is the better fitting function than Eq. \ref{eq_Omega_n} (blue) for the raw data in black when $\Delta\omega_{pu}$ is far-detuned.
Even assuming two TLS groups with $p_1$ = $\sqrt{320}\,p_2$ as shown in green, $\Delta\omega$ is not matching the raw data. 
More details of the two-tone technique are found in the main text.

\section{Uncertainty in the dipole moments}\label{COMSOL}
\setcounter{figure}{0}   
\begin{figure}
    \centering
    \includegraphics[width = 6.5cm]{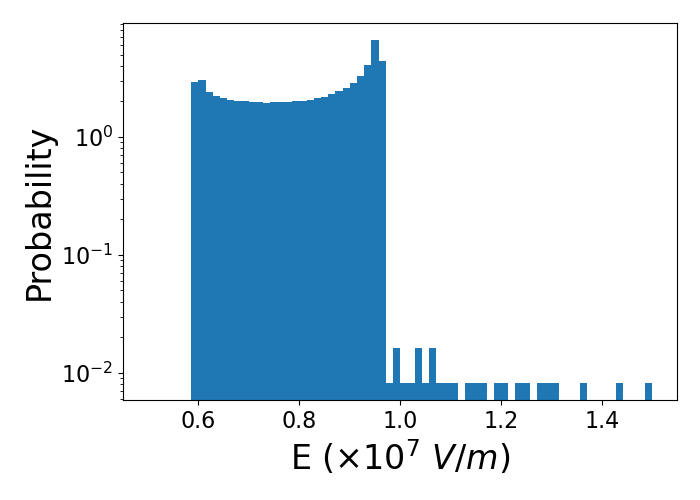}
    \caption{The electric field distribution is simulated by COMSOL with one voltage on one side. The standard deviation of the electric field is 16 \% of the mean value.}
    \label{sfig_comsol}
\end{figure}
The accuracy of $d$ and $n_{pu}$ heavily affect the extracted value of $p$. To obtain the average $d$, 
we compared measured frequencies to simulated frequencies, where the latter is obtained on the resonator structure simulated in COMSOL with meander inductor and uniform gap capacitor. 
The kinetic inductance in our Al meander is not included in the simulation, but the film thickness (100 nm) is much larger than the penetration depth such that the geometric inductance dominates. 

\red{Recall that the gap in the capacitor is not precisely uniform and also that the cooldown is known to bend the capacitor. However, we know the gap at the supports and also the resonator frequency related to the gap distribution. Using these constraints, we simulate the capacitor using a COMSOL simulation and find the average gap distance of $d$ = 125 nm. In the COMSOL modeling the electric field distributions were found in 2D using a structure similar to the upper panel in Fig. 2 b). We assumed the bending of the top bridge followed a quadratic equation symmetrically, related to the supporting posts, where the height is precisely fabricated. The simulation with correction for the same capacitance as the uniform model gives the following: a center height of 0.104 $\mu m$ and a quadratic coefficient of 8.4e-3 ($\mu m^{-1}$). The calculated gap distribution has a standard deviation of $\Delta d =$ 16 \% from the mean value.}

\red{In our measurement of the device, the limit to precision is caused by the non-uniform gap distance $d$. However, we also used the standard practice of calibrating our input power using room temperature data. This results in some systematic inaccuracy to $p$ which we have not carefully characterized. It is possible that our inaccuracy from the input power calibration is larger than the uncertainty caused by capacitor gap non-uniformity.}

\section{Voltage tunable resonance and nonlinear oscillator}\label{appendix_duff}
\setcounter{figure}{0}   

\begin{figure}
    \centering
    \includegraphics[width = 8cm]{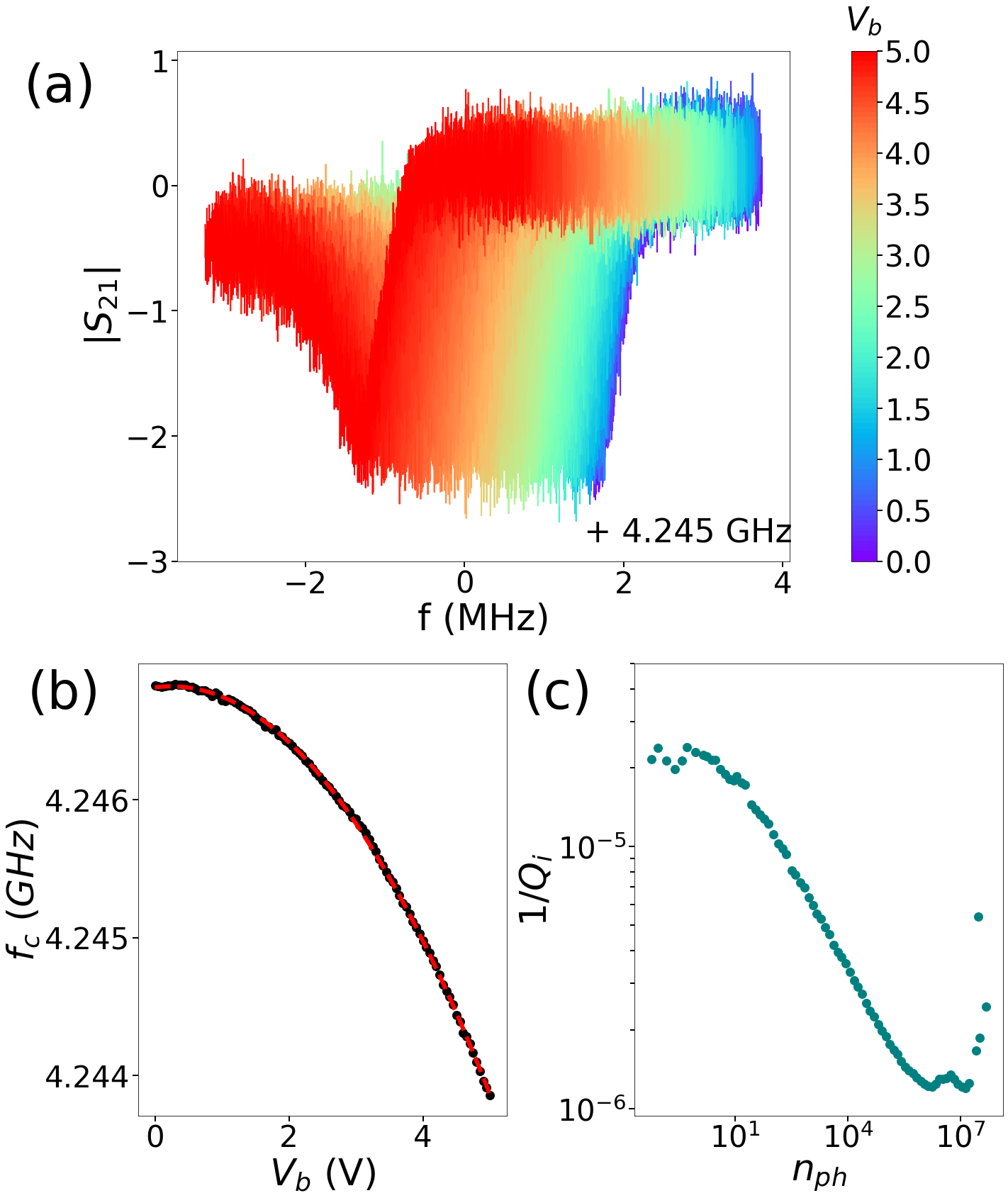}
    \caption{Resonance shift due to voltage biasing. (a) $|S_{21}|$ at different voltage bias $V_b$. Different colors show the $V_b$ from purple for $V_b$ = 0 to red for $V_b$ = 5V. (b) resonance $f_c$ vs. $V_b$. We found a quadratic decrease in $f_c$ and the fit to $m_2V_b^2+m_1V_b+m_0$, shown in the red line, where $m_2$ = -0.13($\frac{MHz}{V^2}$), $m_1$ = 0.06($\frac{MHz}{V}$) and $m_0$= 4.2468(GHz). (c) $1/{Q_i}$ vs. $n_{ph}$ from a SiNx VGC. When $n_{ph}>5\times{10}^6$, $\tan \delta (n_{ph})$ increases due to the high-power-induced quasiparticles. The maximum ac voltage amplitude across the capacitor is $ \sqrt{\frac{2h f_c}{C}n_{ph}} = $ 12 mV, which does not cause an obvious frequency shift or extra complexity, corresponding to $n_{ph}=5\times{10}^6$.}
    \label{sfig_duffing}
\end{figure}
We are able to apply a voltage bias $V_b$ on the capacitors in our design. 
The accumulated charges on the electrodes would attract and bring two electrodes closer resulting in a decrease in the resonance frequency.
In Fig. \ref{sfig_duffing} (a) and (b), we show the magnitude of transmission $S_{21}$ and the resonance $f_c$ at various $V_b$. 
The $f_c(V_b)$ shows a quadratic dependence: $m_2V_b^2+m_1V_b+m_0$, where $m_2$ = -0.13 ($\frac{MHz}{V^2}$), $m_1$ = 0.06 ($\frac{MHz}{V}$) and $m_0$= 4.2468 (GHz).  
The red fit line is shown in Fig. \ref{sfig_duffing} (b).
In Fig. \ref{sfig_duffing} (c), we display the resonator has nonlinearity when stored photon number $n_{ph}>5\times{10}^6$.
The amplitude of the voltage $\delta V = \sqrt{\frac{2h f_c}{C}n_{ph}}$ = 12 mV when $n_{ph}=\ 5\times{10}^6$. 
Since this amplitude is too small to attract the plates, we conclude that the extra loss is due to the quasiparticles created by Cooper pairs broken by up-converted photons.

\section{VGC aging effect}\label{appendix_aging}
\setcounter{figure}{0}   
\begin{figure}
    \centering
    \includegraphics[width = 6cm]{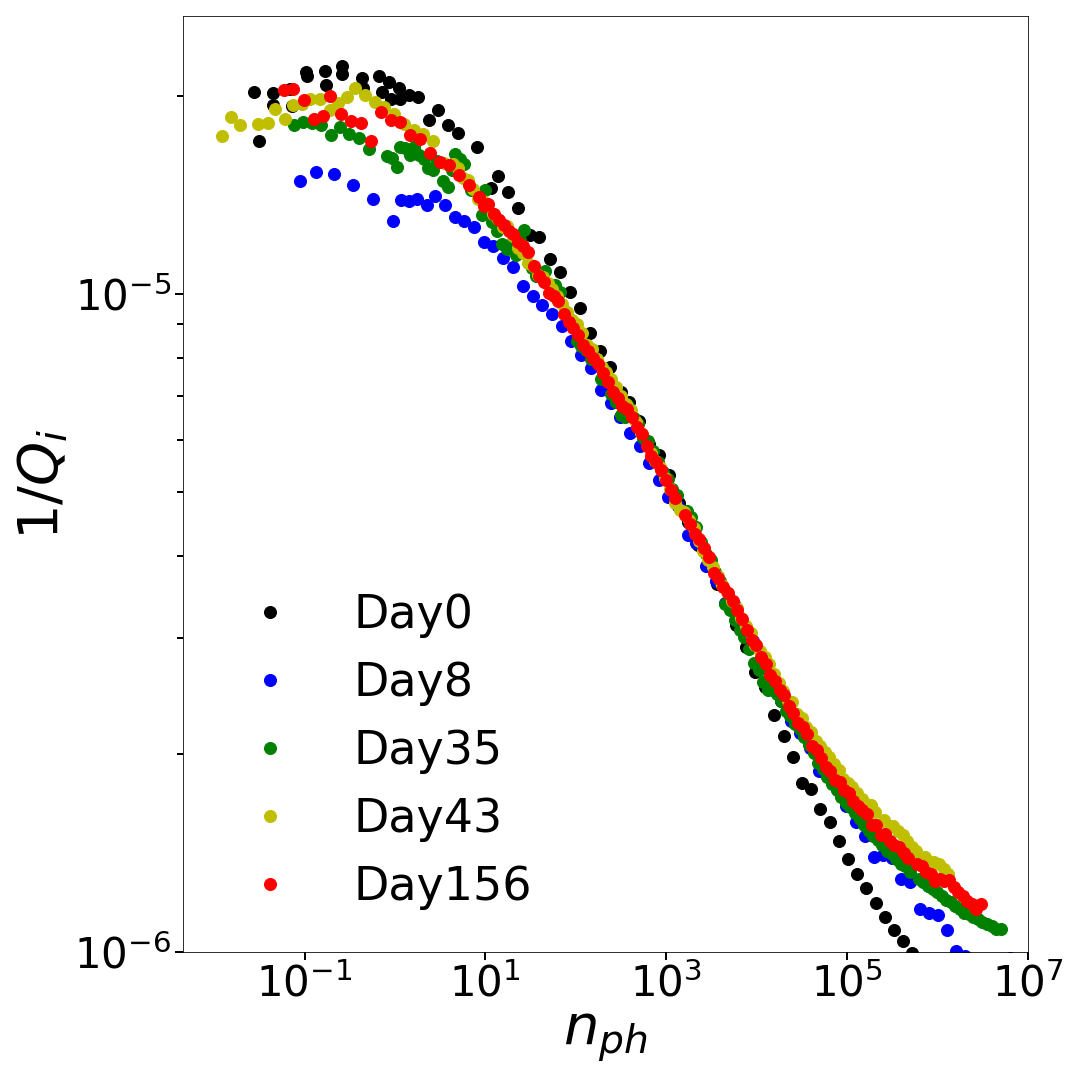}
    \caption{The aging effect on a SiNx VGC. The "Day0" indicates the resonator is in the vacuum right after the releasing process. We find the jitter rate $\gamma_{max}$ is slightly increased with time but the changes are small.}
    \label{sfig_aging}
\end{figure}
We found a small aging effect in our SiNx VGC. In Fig. \ref{sfig_aging}, we showed $1/Q_i$ of one SiNx VGC from right after the VGC was released as Day0 (about 2 hours to be pumped into vacuum) to Day N (where N is the number of days after releasing and of exposure of the MA interface to ambient conditions). 
A slightly increased $\gamma_{max}$ was found over time which implies some accumulation of LF TLSs.

\bibliography{VGR_reference}

\end{document}